\documentclass[aps,prl,preprint,groupedaddress,showpacs]{revtex4}
\usepackage{mathrsfs,bbm} 
\begin{document}
\title{Nonperturbative results for the mass dependence of the ${\rm QED}$ 
fermion determinant}
\author{M. P. Fry}
\affiliation{School of Mathematics, University of Dublin, Trinity College,     Dublin 2, Ireland}
\date{\today}

\begin{abstract}

The fermion determinant in four-dimensional quantum
electrodynamics in the presence of $O(2)\times O(3)$ symmetric background
gauge fields with a nonvanishing global chiral anomaly is
considered. It is shown that the leading mass singularity of the
determinant's nonperturbative part is fixed by the anomaly. It is also
shown that for a large class of such fields there is at least one value
of the fermion mass at which the determinant's nonperturbative part 
reduces to its noninteracting value.

\end{abstract}

\pacs{12.20.Ds, 11.10.Kk, 11.15.Tk}

\maketitle

Every physical process calculable within the standard model
ultimately depends on the model's fermion determinants. These are
part of the effective functional measure for the gauge fields when
the fermion fields are integrated. Without them, charge and color
screening, quark fragmentation into hadrons and unitarity would
be lost. Accordingly, they are fundamental, and the nonperturbative
structure of the standard model requires corresponding information
about its determinants.

Taking Schwinger's 1951 paper \cite{1} on vacuum polarization as the
beginning of the modern era of fermion determinants, progress on
elucidating their nonperturbative properties has been extremely
slow. In this paper we focus on the fermion mass dependence of the determinant in four-dimensional QED.We note here the recent advance in determining
the quark mass dependence of the ${\rm QCD}_4$ instanton determinant \cite{2}.

We begin by summarizing the main analytic results for ${\rm QED}_4$'s
fermion determinant on non-compact, Euclidean space-time.
Formally, a fermion field integration produces the ratio of
determinants ${\rm det}(\lefteqn/p-e\lefteqn\not A+m)/{\rm det}(\lefteqn/p+m)={\rm det}(1-eS\lefteqn\not A)$, where 
$S$ is the free fermion propagator. Since the operator 
$S\lefteqn\not A$ is not trace class, 
${\rm det}(1-eS\lefteqn\not A)$ is undefined 
no matter how well-behaved the gauge field $A_\mu$ is. Nevertheless, 
sense is made of it based on the following results.

$(a)$\  The operator $S\lefteqn\not A$ is a non-Hermitian compact operator in the trace ideal $\mathscr{I}_p$ for $p>4$ and fermion mass $m\ne 0$ provided 
$A_\mu\in L^p(\mathbbm{R}^4)$ \cite{3,4}.
This includes the instanton-like case of $A_\mu$ having a $1/r$ falloff.
The theorem means that the traces ${\rm Tr}(S\lefteqn\not A)^n,\ n\ge 5$, are absolutely convergent and really do correspond to sums of eigenvalues of $S\lefteqn\not A$.

$(b)$\ A renormalized determinant can be defined:

\begin{equation}\label{1}
{\rm det}_{ren}=\exp(\Pi_2+\Pi_3+\Pi_4){\rm det}_5(1-eS\hskip-0,08cm\not\hskip-0,13cm A),
\end{equation}
where

\begin{equation}\label{2}
\ln{\rm det}_5={\rm Tr}\left[\ln(1-eS\hskip-0,08cm\not\hskip-0,13cm A)+\sum_{n=1}^4\frac{(eS\hskip-0,08cm\not\hskip-0,13cm A)^n}{n}\right],
\end{equation}
and $\Pi_{2,3,4}$ are the second, third and fourth-order contributions
to the one-loop effective action, $\ln{\rm det}_{ren}$, defined by some
consistent regularization procedure \cite{5}. The graph $\Pi_2$ contains a
charge renormalization subtraction. The regularization should
result in $\Pi_3=0$ by $C$-invariance and give the unique gauge
invariant result for $\Pi_4$.

$(c)$\ As corollaries of $(a)$ and $(b)$, ${\rm det}_5$ is an entire function of the coupling $e$ and can be represented in terms of the discrete
complex eigenvalues $1/e_n$ of $S\lefteqn\not A$:

\begin{equation}\label{3}
{\rm det}_5=\prod_n\left[\left(1-\frac{e}{e_n}\right)\exp\left(\sum_{k=1}^4\frac{(e/e_n)^k}{k}\right)\right].
\end{equation}

By $C$-invariance and the reality of ${\rm det}_5$ for real $e$ the eigenvalues
can appear in quartets $\pm e_n$, $\pm\bar e_n$ or as imaginary pairs \cite{6}.

$(d)$\ ${\rm det}_{ren}$  has no zeros for real $e$ when $m\ne 0$ \cite{7}, and since ${\rm det}_{ren}(e=0)=1,\ {\rm det}_{ren}>0$ for real $e$.

$(e)$\ ${\rm det}_{ren}$ is an enire function of $e$ of order $4$ since $S\lefteqn\not A \in {\mathscr I}_{4+\epsilon}$ when
$A_\mu\in L^{4+\epsilon}(\mathbbm{R}^4),\ \epsilon>0$ \cite{8}. This conclusion was first reached for a restricted 
class of gauge fields by Adler \cite{9} and later by Balian et al. \cite{10},
for another restricted class of fields. The growth of ${\rm det}_{ren}$ for
real values of $e$ is unknown.

$(f)$\ ${\rm det}_5$ is an analytic function of $m$ throughout the complex
$m$-plane cut along the negative real axis \cite{ 11}.

This comprises the general analytic knowledge of ${\rm det}_{ren}$ obtained
since 1951. The present sparse knowledge of such a central part of the standard model is noteworthy. There is a large body of results
for background gauge fields which do not fall off sufficiently rapidly
in all directions in $\mathbbm{R}^4$ to satisfy the theorem in $(a)$ 
\cite{17} . These require the introduction of an ${\it ad}\ {\it hoc}$ 
volume cutoff, and none of the
results $(b)-(f)$ necessarily hold for such fields.

We report here on an extension of results $(a)-(f)$ for a large
class of $O(2)\times O(3)$ symmetric background gauge fields of the form
$A_\mu(x)=M_{\mu\nu}x_\nu a(r^2)$, where the profile function $a(r^2)$ is at least three times differentiable, regular at the origin, and $a(r^2)=\nu/r^2$ for $r>R$, $R$ being a range parameter, $r^2=x_\mu x_\mu$. The constant $\nu$ is assumed positive without loss of generality. For $r<R$, $a(r^2)$ may have multiple zeros. The constant antisymmetric matrix $M$ has nonvanishing entries $M_{12}=M_{30}=1$. Letting $ ^*F_{\mu\nu}=\frac12\epsilon_{\mu\nu\alpha\beta}F_{\alpha\beta}$ and noting that $ ^*F\, F=\partial_\alpha(\epsilon_{\alpha\beta\mu\nu}A_\beta F_{\mu\nu})$, it is evident that $A_\mu$  must have a $1/r$ falloff for the Dirac operator $\lefteqn\not D=\lefteqn/P-e\lefteqn\not A$ to have a nonvanishing global chiral anomaly ${\cal A}=-\int\,d^4x\,^*F_{\mu\nu}F_{\mu\nu}/16\pi^2$. In our case ${\cal A}=\nu^2/2$. From here on we set $e=1$ to reduce notation.

In \cite{11} we proved a vanishing theorem for this choice of $A_\mu$.
Consequently all the square-integrable zero modes of $\lefteqn\not D$ have
positive chirality, and such modes first appear when $\nu>2$. The zero modes can be shifted to the negative chirality sector by replacing $M$ with the antisymmetric matrix $N$ with entries $N_{03}=N_{12}=1$. Since $F_{\mu\nu}$ is not self-dual it extends the vanishing theorem of \cite{12} to such U(1) fields. Because the calculation here is in non-compact Euclidean space-time the index
theorem has to be modified to account for the continuum part of the
spectrum of $\lefteqn\not D^2$ extending down to zero energy, thereby contributing an additional part to the index of $\lefteqn\not D$ \cite{13,14}. These low energy states play an essential role in the analysis discussed below.

We can now add the following results to the list $(a)-(f)$ above:

$(g)$\ For the class of gauge fields defined above the leading mass singularity of $\ln{\rm det}_5$ is governed by the chiral anomaly, that is,

\begin{equation}\label{4}
{\ln{\rm det}_5}{{{\phantom{qwery}}\atop{\widetilde{\scriptstyle m\to 0}}}}\frac{\nu^2}{4}\ln\,m^2+{\rm less\ singular},
\end{equation} 
so that $\ln{\rm det}_5$ becomes negative as $m\to 0$. Inconclusive evidence for $(4)$ was first reported in \cite{11}.

The presence of a zero mode in the spectrum of the Dirac operator and its control of the leading quark mass dependence of ${\rm QCD}_4$ instanton determinant has been known for many years \cite{15}.
Establishing (\ref{4}) relies on the above vanishing theorem and showing that the zero-mode-free negative chirality sector contributes terms to (\ref{4}) less singular than $\ln\,m^2$. With more effort it should be possible to prove (\ref{4}) under the more general assumption that $a(r^2)\sim\nu/r^2$, for $r\gg R$.

$(h)$\ If

\begin{equation}\label{5}
\begin{array}{rr}
\displaystyle\int\limits_0^{R^2}dr^2[2r^{14}a^{'6}+12r^{12}aa^{'5}+23r^{10}a^2a^{'4}+ \\
12r^8a^3a^{'3}-19r^6a^4a^{'2}]<\displaystyle\frac{9\nu^6}{2R^8},
\end{array}
\end{equation}
then $\ln{\rm det}_5$ becomes positive before dropping off to zero for $m\to\infty$ \cite{11}. This and (\ref{4}) imply that there is at least one value of $m$ for which $\ln{\rm det}_5=0$. That is, $\ln{\rm det}_5$ has a mass zero, or possibly an odd number of such zeros, at which ${\rm det}_5=1$.

Subject to the conditions on $a$ stated above, (\ref{5}) can be satisfied in general for an $a(r^2)$ that can oscillate between positive and negative values before curving downward to join smoothly with its long-range form $\nu/r^2$ at $r=R$.

As $\nu$ is varied the eigenvalues $e_n$ in (\ref{3}) will shift, presumably
shifting the position of the mass zero as well. That a particular value
of $m$ can cause ${\rm det}_5$ to assume its noninteracting value indicates that mass has a profound effect on the distribution of the eigenvalues $e_n$ in the
complex plane. 

To establish (\ref{4}) we begin by defining ${\rm det}_{ren}$ in (\ref{1}) by

\begin{equation}\label{6}
\begin{array}{cc}
\ln{\rm det}_{ren}=\frac{1}{2}\lim\limits_{\delta\to 0}\displaystyle\int\limits_\delta^\infty\frac{dt}{t}\left[{\rm Tr}\left(e^{-tP^2}-e^{-t[(P-A)^2+\frac12\sigma F]}\right)\right.  \\
\left.\times\,e^{-tm^2}+\frac{1}{24\pi^2}\displaystyle\int\,d^4xF^2(x)e^{-t\mu^2}\right],\end{array}
\end{equation}
where $\mu$ is the renormalization scale. Although ${\rm det}_{ren}$ is finite
with on-shell renormalization of ${\Pi}_2$ when $F_{\mu\nu}\sim 1/r^2$, this complicates the small mass analysis of ${\rm det}_5$, and so we prefer to deal with the off-shell case. One can always go back on-shell once ${\rm det}_5$ is
understood.

In the representation where $\gamma_5$ is diagonal with entries $\pm\mathbbm{1}_2$, $(P-A)^2+\frac12\sigma F$ is diagonal with corresponding positive and negative chirality entries $H_{\pm}=(P-A)^2-{\sigma\hskip-0,23cm\sigma}\cdot{\bf(B\pm E)}$. Differentiating (\ref{6}) with respect to $m^2$ yields the renormalization independent result

\begin{equation}\label{7}
\begin{array}{ccc}
\displaystyle m^2\frac{\partial}{\partial m^2}\ln{\rm det}_5=\textstyle\frac12 m^2{\rm Tr}\left[(H_++m^2)^{-1}-(H_-+m^2)^{-1}\right] \\
+m^2\displaystyle\int\limits_0^\infty dt e^{-tm^2}\displaystyle\int d^4x{\rm tr}<x|e^{-tH_-}-e^{-t{P}^2}|x> \\
\hskip-2,7cm -m^2\partial\Pi_2/\partial m^2-m^2\partial\Pi_4/\partial m^2,
\end{array}
\end{equation}
where we have taken $\delta$ to zero and set $\Pi_3=0$. The first and last terms in (\ref{7}) are well-defined for the background fields considered here,
but the second and third terms are not. Specifically, the perturbative
expansion of (\ref{6}) gives $\partial{\Pi}_2/\partial m^2\sim\int\limits_0 dk/k$ which must be cancelled.

The strategy is this: There must be a corresponding infrared
divergence in the second term in (\ref{7}) that cancels that in $\partial\Pi_2/\partial m^2$, as the left-hand side of (\ref{7}) is well-defined for $m^2>0$. The second term in (\ref{7}) will be calculated by summing over the exact eigenstates
of $H_-$ . As already noted, these are scattering states only. An
infrared regulator is introduced by cutting off the low energy
spectra of $H_-$ and $P^2$ at $\lambda^2$. Then the infrared divergent part is
isolated; it must be second order. Identify it as the divergent
part of $\partial\Pi_2/\partial m^2$ to effect the cancellation of infrared divergences. That is, the second term in (\ref{7}) defines the divergent part of
$\partial\Pi_2/\partial m^2$ according to (\ref{9}) and (\ref{13}) below, consistent with our way of calculating ${\rm det}_{ren}$. Then set $\lambda=0$ and finally study the $m\to 0$ limit of (\ref{7}) to find the small mass dependence of the well-defined quantity $\ln\,{\rm det}_5$. It is essential to proceed in this way. Pulling out the contribution
to $\partial\Pi_2/\partial m^2$ from the second term in (\ref{7}) by a straightforward perturbation expansion results in a gauge invariant remainder that
is a sum of separately non-gauge invariant terms, leading to a
computational impasse.

For the fields under consideration we find $\Pi_4$ is well-defined
for $m\ne 0$ and is less singular than $\ln\,m^2$ as $m\to 0$ and so gives a
vanishing contribution to the right-hand side of (\ref{7}) as $m\to 0$.
This result relies in part on the finiteness at $m=0$ of
the photon-photon scattering subgraph in $\Pi_4$ \cite{16}.

We now turn to the calculation of the rest of the right-hand side
of (\ref{7}). Here we use the definition of the chiral anomaly on non-compact
manifolds due to Musto et al. \cite{14}, ${\cal A}=\lim\limits_{m\to 0}\,m^2{\rm Tr}\,[(H_++m^2)^{-1}-(H_-+m^2)^{-1}]$.
This combined with ${\cal A}=\nu^2/2$, Eq.(\ref{7}), and an integration with respect to $m^2$ gives result (\ref{4}), provided that the remainder in (\ref{7}) contributes terms to (\ref{4}) less singular than $\ln\,m^2$ for $m\to 0$.

Denote the second term on the right-hand side of (\ref{7}) by I and
obtain for $m\to 0$

\begin{equation}\label{8}
\begin{array}{rr}
I(m^2)=m^2\lim\limits_{\lambda\to 0}\lim\limits_{L\to\infty}\displaystyle\int\limits_{\lambda^2}^{\Lambda^2}\frac{dk^2}{k^2+m^2}\displaystyle\int\limits_0^Ldr\,r^3\displaystyle\int d\Omega_4 \\
\times\sum\limits_{jMmm'}\left(|\Psi_{EjMmm'}^-(x)|^2-|\Psi_{EjMmm'}^0(x)|^2\right),
\end{array}
\end{equation}
where $\Psi_{EjMmm'}^-$ are the eigenstates $H_-$ derived in \cite{11}, and
$\Psi_{EjMmm'}^0$ are the associated free-particle states. Here $E=k^2$, and
$j=0,\frac12,1,..;\,M=-j-\frac12,..,j+\frac12;\,m=-j,..,j;\,m'=\pm\frac12$ are the quantum numbers associated with the $O(2)\times O(3)$ symmetry of the background fields; $\lambda$ is the infrared cutoff introduced above, and $\Lambda$, with $\Lambda R<1$, limits the range of $k$ needed to study the small-mass dependence of $I$.

Divide $I$ into $I_>(I_<)$, the exterior (interior) parts of $I$ from $r>R$
$(r<R)$, and consider first the most singular part in $m^2,\,I_>$.
The radial wave functions associated with $\Psi^-$ for $r>R$ are calculated
from the outgoing wave combination of Bessel functions
$\sqrt rJ_\sigma(kr)\cos\Delta_\alpha(k)-\sqrt rY_\sigma(kr)\sin\Delta_\alpha(k)$, where $\alpha$ denotes $jMm'$, 
$\sigma=\sqrt{(2j+1)^2+4\nu M+\nu^2}$, and $\Delta_\alpha$ is the energy-dependent part of the low-energy phase shifts 
$\delta_{\alpha}$, $\Delta_\alpha(k)=\pi(\sigma-2j-1)/2+\delta_\alpha(k)$, mod$\,\pi$. Since $\Delta_\alpha(0)=0$, we can expand in powers of $\Delta_\alpha$. For $|M|\ne j+\frac12,\ {\rm tan}\Delta_\alpha=C_\alpha(\sigma)(kR/2)^{2\sigma}(\sigma\Gamma^2(\sigma))^{-1}\left[1+O((kR)^2,(kR)^{2\sigma})\right]$, where $C_\alpha(\sigma)$
is a bounded function of $\sigma$ \cite{11}. The rapid falloff
of $\Delta_\alpha$ with $j$ and energy allows one to terminate the expansion after $\Delta_\alpha^2$. Terms in $I_>$ containing $\Delta_\alpha$ and $\Delta_\alpha^2$ are uniformly convergent and can be integrated term-by-term and the limit $L=\infty$ taken. There are some oscillating $k$-integrals containing $\cos(2kL)$ and $\sin(2kL)$. These are set equal to zero by the Riemann-Lebesque
lemma following the sequence of limits in (\ref{8}). The result is
contributions to $I_>$ less singular than $m^2\ln\,m^2$ and $O(1)$ contributions to $\ln{\rm det}_5$. Terms from $M=j+\frac12$ contribute $O(m^2)$ terms to $I_>$. Terms of $O(\Delta_\alpha^0)$ will be considered below.

The zero modes of $H_+$ appear in the sector $M=-j-\frac12$ for values of
$j$ satisfying $\nu>2j+2,\,j=0,\frac12,..$. The most singular contribution to $I_>$ occurs at the zero mode thresholds $M=-j-\frac12,\,\nu=2j+2$ at which $\Delta_\alpha$'s energy dependence drops to ${\rm tan}\,\Delta_\alpha=\frac{\pi}{2}(1+O(kR)^2)[\ln(kR)+C+O(kR)^2\ln(kR)]^{-1}$
where $C$ is a negative $k$-independent constant \cite{11}. This results in a
contribution to $I_>$ of $O(1/\ln(mR))$ and a $\ln\left|\ln(mR)\right|$ contribution to
$\ln{\rm det}_5$ in (\ref{4}). This covers all terms in (\ref{8}) from $\Delta_\alpha$ and $\Delta_\alpha^2$.

The zero mode thresholds also dominate the region $r<R$. Specifically,
they are responsible for the radial wave function contributing to (\ref{8})
with the slowest $k$ falloff, whose form is $(\ln kR+C)^{-1}\psi(k^2,r)$,
$\psi(0,r)\ne 0$, and $C$ as above. Here $\psi$ is analytic in $k^2$
and is a smooth function of $r$ behaving near $r=0$ as $r^{2j+3/2}$. This
results in contributions of $O(m^2)$ to $I_<$ and $O(1)$ to $\ln{\rm det}_5$. Other cases in the $M=-j-\frac12$ sector have a faster small $k$ falloff. The study of the $M\ne -j-\frac12$ sectors is facilitated by the $1/(2j+1)!$ falloff of the radial wave functions (also true for $M=-j-\frac12$), their small $k$ falloff of at least $(kR)^\sigma$, and their $r^{2j+1/2}$ behavior near $r=0$ \cite{11}. These results allow the $m\to 0$ limit of $I_<$ to be taken term-by-term, giving
a final $O(1)$ contribution to $\ln{\rm det}_5$.

Now consider the terms in $I_>$ of $O(\Delta_\alpha^0)$, here denoted by $I_>^0$. For fixed $L$ the integral and sum over $j$ in (\ref{8}) can be interchanged
since $|J_\sigma(z)|\le|z/2|^\sigma/\Gamma(\sigma+1)$, $z$ real, and because only $J_\sigma$ is present in $I_>^0$. The result for the $L$-dependent terms is
\begin{equation}\label{9}
I_>^0=\frac{m^2}{4}\lim\limits_{\lambda\to 0}\lim\limits_{L\to\infty}\int\limits_{\lambda^2}^{\Lambda^2}\frac{dk^2}{k^2+m^2}L^2\Bigl(S_1(kL)+S_2(kL)\Bigr),
\end{equation}
where
\begin{equation}\label{10}
\begin{array}{cc}
S_1=\displaystyle\sum_{j=0,\frac12,..}(2j+1)\Bigl[J_{|2j+2-\nu|}^2-J_{|2j+1-\nu|}J_{|2j+3-\nu|}\Bigr.\\
\hskip2cm \Bigl.+(\nu\to -\nu)-(\nu=0)\Bigr],
\end{array}
\end{equation}
\begin{equation}\label{11}
\begin{array}{cc} 
S_2=\sum\limits_{j=\frac12,1,..}(2j+1)\sum\limits_{M=-j+\frac12}^{j-\frac12}\Bigl[J_{\sigma+1}^2-J_{\sigma+2}J_\sigma+J_{\sigma-1}^2\Bigr. \\
\hskip4cm \Bigl. -J_{\sigma-2}J_\sigma-(\nu=0)\Bigr].
\end{array}
\end{equation}
The Bessel functions are evaluated at $kL$. These series are not
uniformly convergent in $L$ and must be summed before taking $L\to\infty$.
$S_1$ can be summed to give for $kL\gg 1$
\begin{equation}\label{12}
S_1=\frac{2\nu^2}{\pi kL}+\frac{1}{\pi(kL)^2}\cos(2kL)\sin^2\left(\frac{\pi\nu}{2}\right)+O(kL)^{-3}.
\end{equation}
The leading term in (\ref{12}) must be cancelled by $S_2$ to make $I_>^0$ finite. Because of the absolute value signs in (\ref{10}), Eq.(\ref{12}) is valid for $\nu\le 1$. For $\nu>1$ we show in Appendix A that the leading term in (\ref{12}) is unchanged.

We have not been able to sum $S_2$  without resorting to its 
expansion in $\nu$. This is a well-behaved expansion as it occurs in
the Bessel function's order, and $J_\sigma$ is an entire function of $\sigma$.
To $O(\nu^2)$ we find for $kL\gg 1$
\begin{equation}\label{13}
\begin{array}{rr}
S_2=-\displaystyle\frac{2\nu^2}{\pi kL}+\left(\frac{\nu}{kL}\right)^2\left[C+\frac{\pi}{12}\cos(2kL)\right. \\
\ \\
\left.+\left(\displaystyle\frac{7}{30}+\frac{2}{3}\ln 2\right)\sin(2kL)\right]+O(kL)^{-3}.
\end{array}
\end{equation}
As expected, the leading term in (\ref{13}) cancels that in (\ref{12}). Referring to (\ref{9}), the second term in (\ref{13}) results in the expected infrared
divergent term discussed above. The constant $C$ is given by a
complicated, but absolutely convergent, series of Bessel functions.
Its value is irrelevant to our analysis as it will be cancelled
by the counterterm $\partial\Pi_2/\partial m^2$. The remaining oscillating terms in (\ref{12}) and (\ref{13}) give vanishing contribution to $I_>^0$ by the
Riemann-Lebesque lemma.

The second-order calculation of $S_2$ may be extended to all orders
in $\nu$. Structures generated in second order appear again in higher
orders differentiated with respect to Bessel function order. No
further infrared divergences appear, only $\cos(2kL)$ and $\sin(2kL)$
terms as in $(\ref{12})$ and $(\ref{13})$. Appendices A and B show that (\ref{13}) and higher order results for $S_2$ hold for $\nu\le 2$ and that they can be continued to $\nu>2$. There is no loss of information in the expansion of $S_2$ and its continuation.
The $R$-dependent terms from the lower bound of integration of $I_>^0$ are uniformly convergent, and no expansion is necessary. They result in $O(1)$
contributions to $\ln{\rm det}_5$. This establishes Eq.$(4)$.

The conclusion that ${\rm det}_5$ can be reduced to its noninteracting value by varying its mass for a class of background gauge fields points to an unexpected nonperturbative role of mass in ${\rm QED}_4$'s effective action. It would be surprising if result $(4)$ -- the chiral anomaly's control of $\ln{\rm det}_5$'s leading mass singularity -- is limited to our background fields. Presumably it is generally true and, if so, mass zero(s) in $\ln{\rm det}_5$ are also present more generally.

\begin{acknowledgements}
The communication of Christian Schubert that $\Pi_4$ in scalar $QED_4$ with a $r^{-2}$ falloff profile function has no $\ln\,m^2$ singularity is gratefully acknowledged.
\end{acknowledgements}

\newpage

\begin{center}
 APPENDIX A
\end{center}

\setcounter{equation}{0}
\renewcommand{\theequation}{A\arabic{equation}}

To calculate $S_1(\nu)$ first assume $\nu\le 1$ so that the absolute
value signs can be dropped. Combine products of Bessel functions
using the identity \cite{18}

\begin{equation}\label{A1}
J_\mu(z)J_\nu(z)=\frac2\pi\int\limits_0^{\pi/2}\,d\theta J_{\mu+\nu}(2z\cos\theta)\cos[(\mu-\nu)\theta],
\end{equation}
for ${\rm Re}(\mu+\nu)>-1$. Then calculate sums using

\begin{equation}\label{A2}
\sum\limits_{n=0}^{\infty}(2n+\rho)J_{2n+\rho}(z)=\frac{z}{2}J_{\rho-1}(z)+\frac{z}{2}\int\limits_0^z\,dtJ_\rho(t),
\end{equation}
for ${\rm Re}\,\rho>-1$ from \cite{19}, entry 5.7.2.6 and

\begin{equation}\label{A3}
\sum\limits_{n=0}^\infty J_{2n+\rho}(z)=\frac12\int\limits_0^z\,dtJ_{\rho-1}(t),\end{equation}
from entry 5.7.1.6. Calculation of standard integrals followed by the
limit $z=kL\gg 1$ gives Eq.(\ref{12}).

When $\nu>1$ care has to be taken in summing over $j$ and the sign
of the absolute values in Eq.(\ref{10}). Let $\nu=N+\epsilon,\ 0<\epsilon\le 1,\ N=1,2,\ldots$. Then the absolute value signs can be omitted for $j\ge N/2$. The sum in (\ref{10}) is the same as in (\ref{12}) with the addition of the extra term
$$
(-1)^N\left[\frac{N+1}{2}\right]\frac{\cos(2kL)\cos(\pi\nu)}{2\pi(kL)^2}+O(kL)^{-3},
$$
where $[x]$ is the integral part of $x$.

Now consider $S_2$ in Eq.(\ref{11}). From (\ref{A1}) we get

$$
\begin{array}{cc}
J_{\sigma+1}^2-J_{\sigma+2}J_{\sigma}+J_{\sigma-1}^2-J_{\sigma-2}J_{\sigma} \\
=\displaystyle\frac4\pi\int\limits_0^{\pi/2}\,d\theta\sin^2\theta\left[J_{2\sigma+2}(2z\cos\theta)+J_{2\sigma-2}(2z\cos\theta)\right].
\end{array}
$$
Expand $\sigma$ in powers of $\nu$, initially assuming $\nu\le 2$ so that the
expansion of $\sigma$ converges for $j\ge 1/2$. Keeping all terms through
order $\nu^2$ we get after summing over $M$,

\begin{equation}\label{A4}
\begin{array}{cc}
S_2=\displaystyle\frac{8\nu^2}{3\pi}\frac{\partial}{\partial\rho}\int\limits_0^1dx\sqrt{1-x^2}\sum\limits_{j=1/2,1,..}\left(2j+1-\frac{1}{2j+1}\right)J_{4j+4+\rho}(2xz)\Biggl|_{\rho=0} \\
\hskip-0,83cm+\displaystyle\frac{16\nu^2}{3\pi}\frac{\partial^2}{\partial\rho^2}\int\limits_0^1dx\sqrt{1-x^2}\sum\limits_{j=1/2,1,..}(2j^2-1)J_{4j+4+\rho}(2xz)\Biggr|_{\rho=0}. \\
\end{array}
\end{equation}
In addition to (\ref{A2}) and (\ref{A3}) use is made of the following results
from \cite{19}:

\begin{equation}\label{A5}
\sum\limits_{n=0}^\infty(2n+\rho)^2J_{2n+\rho}(z)=\frac{\rho z}{2}J_{\rho-1}(z)+\frac{z^2}{2}J_\rho(z)+\frac{z^2}{2}\int\limits_0^zdtJ_{\rho+1}(t),
\end{equation}
from entry 5.7.2.24 and

\begin{equation}\label{A6}
\sum\limits_{n=0}^\infty\frac{1}{2n+1}J_{2n+\rho}(z)=\frac{\pi(\rho-1)}{4}\int\limits_0^zdt\hskip0,07cm\frac1t\hskip0,07cm J_{\rho-1}(t){\bf H_0}(z-t),
\end{equation}
for ${\rm Re}\,\rho>1$ from entry 5.7.3.5, where ${\bf H_0}$ is the Struve function of order zero. Then (\ref{A4}) together with (\ref{A2}), (\ref{A3}), (\ref{A5}), (\ref{A6}) becomes

\begin{equation}\label{A7}
\begin{array}{lllll}
S_2=\displaystyle\frac{8\nu^2}{3\pi}\frac{\partial}{\partial\rho}\int\limits_0^1dx\sqrt{1-x^2}\left[\left\{\int\limits_0^{2xz}dt\left[\frac12xzJ_{4+\rho}(t)-\left(\frac12+\frac{\rho}{4}\right)J_{3+\rho}(t)\right]\right.\right. \\
\left.\Biggl.+\displaystyle\frac12xzJ_{3+\rho}(2xz)+\sum\limits_{n=1}^\infty\frac{J_{2n+4+\rho}(2xz)}{(2n+1)(n+1)}\Biggr\}+\left\{\rho\to\rho-4\right\}\right]_{\rho=0} \\
+\displaystyle\frac{8\nu^2}{3\pi}\frac{\partial}{\partial\rho}\int\limits_0^1dx\sqrt{1-x^2}\left[-\pi(\rho+3)\int\limits_0^{2xz}dt\frac1tJ_{3+\rho}(t){\bf H_0}(2xz-t)+J_{4+\rho}(2xz)\right. \\
\displaystyle-J_\rho(2xz)-\frac23J_{\rho+2}(2xz)\Biggl.+8\sum\limits_{n=0}^{\infty}\frac{J_{2n+4+\rho}(2xz)}{(2n+1)(2n+5)}\Biggr]_{\rho=0}\hskip-0,2cm+\frac{16\nu^2}{3\pi}\frac{\partial^2}{\partial\rho^2}\int\limits_0^1dx\sqrt{1-x^2} \\
\times\displaystyle\left[\left\{\int\limits_0^{2xz}dt\left[\left(\frac14(xz)^2+\frac32+\frac{5\rho}{8}+\frac{\rho^2}{16}\right)J_{3+\rho}(t)-\frac14(5+\rho)xzJ_{4+\rho}(t)\right]\right.\right. \\
\displaystyle-\frac14(xz)^2J_{4+\rho}(2xz)\Biggl.-\left(\frac34+\frac18\rho\right)xzJ_{3+\rho}(xz)\Biggl.\Biggr\}+\left\{\rho\to\rho-4\right\}\Biggr]_{\rho=0}\hskip-0,1cm+O(\nu^4). 
\end{array}
\end{equation}

The only terms in (\ref{A7}) needing further discussion are the
infinite series (see below) and

\begin{equation}\label{A8}
I(z)\equiv\int\limits_0^1dx\sqrt{1-x^2}\int\limits_0^{2xz}dt\hskip0,1cm\frac1t\hskip0,1cmJ_{3+\rho}(t){\bf H_0}(2xz-t)=I_1+I_2,
\end{equation}
where

\begin{equation}\label{A9}
\begin{array}{lll}
I_1=\displaystyle\int\limits_0^1dx\sqrt{1-x^2}\int\limits_0^\infty dt\,\frac1t\,J_{3+\rho}(t){\bf H_0}(2xz-t) \\
\hskip0,5cm=\displaystyle\int\limits_0^1dx\sqrt{1-x^2}\int\limits_0^Ndt\,\frac1t\,J_{3+\rho}(t){\bf H_0}(2xz-t)\,(\equiv I_{11}) \\
\hskip0,6cm+\displaystyle\int\limits_0^1dx\sqrt{1-x^2}\int\limits_N^\infty dt\,\frac1t\,J_{3+\rho}(t){\bf H_0}(2xz-t)\,(\equiv I_{12}). \\
\end{array}
\end{equation}
Since for $z\gg 1$ \cite{18}

\begin{equation}\label{A10}
{\bf H_0}(z)\sim Y_o(z)+\frac2\pi\left(\frac1z-\frac{1}{z^3}+\ldots\right),
\end{equation}
then $\lim\limits_{z\to\infty}I_{11}=0$. Since $|{\bf H_0}(z)|<1$ for all $z\ge 0$,

\begin{equation}\label{A11}
|I_{12}|\le\int\limits_0^1dx\sqrt{1-x^2}\int\limits_N^\infty dt\,\frac1t\,|J_{3+\rho}(t)|<\infty,
\end{equation}
Hence, $I_{12}$ concerges uniformly for all $z>0$. Therefore, the limit
$z\to\infty$ can be interchanged with the integrals defining $I_{12}$ to give
$\lim\limits_{z\to\infty}I_{12}=0$, and so $\lim\limits_{z\to\infty}I_1=0$. By inspection

\begin{equation}\label{A12}
I_2=-\int\limits_0^1dx\sqrt{1-x^2}\int\limits_{2xz}^{\infty}dt\,\frac1t\,J_{3+\rho}(t){\bf H_0}(2xz-t),
\end{equation}
vanishes as $z\to\infty$. Hence, $\lim\limits_{z\to\infty}I(z)=0$.

From \cite{18} ${\bf H_0}=2/\pi-{\bf H_1}$ so that

\begin{equation}\label{A13}
\frac{dI}{dz}=2\int\limits_0^1dx\sqrt{1-x^2}\int\limits_0^{2xz}dt\,\frac1t\,J_{3+\rho}(t)\left[\frac2\pi-{\bf H_1}(2xz-t)\right],
\end{equation}
since ${\bf H_0}(0)=0$. Substitute \cite{20}

\begin{equation}\label{A14}
{\bf H_1}(z)=\frac2\pi-\frac2\pi J_o(z)+\frac4\pi\sum\limits_{n=1}^{\infty}\frac{J_{2n}(z)}{4n^2-1},
\end{equation}
in (\ref{A13}) and use \cite{18} for ${\rm Re}(\mu)>0,\, {\rm Re}(\nu)>-1$,

\begin{equation}\label{A15}
\int\limits_0^zdt\,\frac1t\,J_\mu(t)J_\nu(z-t)=\frac{J_{\mu+\nu}(z)}{\mu},
\end{equation}
to get

\begin{equation}\label{A16}
\frac{dI}{dz}=\frac{4/\pi}{3+\rho}\int\limits_0^1dx\,x\sqrt{1-x^2}J_{3+\rho}(2xz)+I_3,
\end{equation}
where

\begin{equation}\label{A17}
I_3=-\frac8\pi\int\limits_0^1dx\,x\sqrt{1-x^2}\int\limits_0^{2xz}dt\,\frac1t\,J_{3+\rho}(t)\sum\limits_{n=1}^{\infty}\frac{J_{2n}(2xz-t)}{4n^2-1}.
\end{equation}
The series in (\ref{A17}) for $0\le 2xz-t\le 2z$ is uniformly convergent by
inspection using $|J_\mu(z)|\le 1,\,\mu\ge 0$. Therefore, the series defines a
continuous function of $2xz-t$, and $J_{3+\rho}(t)/t$ is continuous for
$0\le t\le 2xz$. Hence the series can be integrated term-by-term with
respect to $t$. Using (\ref{A15}),

\begin{equation}\label{A18}
I_3=-\frac{8/\pi}{3+\rho}\int\limits_0^1dx\,x\sqrt{1-x^2}\sum\limits_{n=1}^{\infty}\frac{J_{2n+3+\rho}(2xz)}{4n^2-1}.
\end{equation}
By the same reasoning as above the integral and sum in (\ref{A18}) can
be interchanged to give, with (\ref{A1}) employed in both (\ref{A16}) and (\ref{A18}),

\begin{equation}\label{A19}
\begin{array}{cc}
\displaystyle\frac{dI}{dz}=\frac{1/2}{3+\rho}\left[J_{2+\rho/2}(z)J_{1+\rho/2}(z)-J_{3+\rho/2}(z)J_{\rho/2}(z)\right] \\
\hskip2,7cm-\displaystyle\frac{1}{3+\rho}\sum\limits_{n=1}^{\infty}\frac{J_{n+2+\rho/2}(z)J_{n+1+\rho/2}(z)-J_{n+3+\rho/2}(z)J_{n+\rho/2}(z)}{4n^2-1}. 
\end{array}
\end{equation}
From (\ref{A7}) we need $\partial\left[(\rho+3)I\right]/\partial\rho\biggr|_{\rho=0}$ for large $z\,(=kL)$. By inspection
the series in (\ref{A19}) is uniformly convergent in $z$, allowing the
interchange of limit and sum. Keeping terms of order $z^{-3}$ we get

\begin{equation}\label{A20}
\frac{\partial}{\partial\rho}(\rho+3)\left.\frac{dI}{dz}\right|_{\rho=0}=-\frac{\pi}{4}\frac{\cos(2z)}{z^2}+O\left(\frac{\cos(2z)}{z^3},\,\frac{\sin(2z)}{z^3}\right),
\end{equation}
where we used

\begin{equation}\label{A21}
\sum\limits_{n=1}^{\infty}\frac{1}{4n^2-1}=\frac12,\quad \sum\limits_{n=1}^{\infty}\frac{(-1)^n}{4n^2-1}=\frac12-\frac{\pi}{4}.
\end{equation}
The remainder in (\ref{A20}) has been verified to be finite. From our
previous result $\lim\limits_{z\to\infty}I=0$ we get

\begin{equation}\label{A22}
\frac{\partial}{\partial\rho}(\rho+3)\int\limits_0^1dx\sqrt{1-x^2}\int\limits_0^{2xz}\hskip-0,07cmdt\,\frac1t\,J_{3+\rho}(t){\bf H_0}(2xz-t)\Biggr|_{\rho=0}\hskip-0,11cm=-\frac{\pi}{8}\frac{\sin(2z)}{z^2}+O\left(\frac{\cos(2z)}{z^3},\,\frac{\sin(2z)}{z^3}\right),
\end{equation}
as a contribution to (\ref{A7}).

Regarding the constant $C$ in (\ref{13}), refer to (\ref{A7}) and let

\begin{equation}\label{A23}
f(z,\rho)=\int\limits_0^1dx\sqrt{1-x^2}\left[\sum\limits_{n=1}^{\infty}\frac{J_{2n+4+\rho}(2xz)+(\rho\to\rho-4)}{(2n+1)(n+1)}+8\sum\limits_{n=0}^{\infty}\frac{J_{2n+4+\rho}(2xz)}{(2n+1)(2n+5)}\right].
\end{equation}
These series can be integrated term-by-term due to their uniform
convergence in $x$. From (\ref{A1}),

\begin{equation}\label{A24}
\begin{array}{cc}
f(z,\rho)=\displaystyle\frac{\pi}{4}\sum\limits_{n=1}^{\infty}\frac{J_{n+2+\rho/2}^2(z)-J_{n+3+\rho/2}(z)J_{n+1+\rho/2}(z)+(\rho\to\rho-4)}{(2n+1)(n+1)} \\
\hskip-1,15cm+\displaystyle2\pi\sum\limits_{n=0}^{\infty}\frac{J_{n+2+\rho/2}^2(z)-J_{n+3+\rho/2}(z)J_{n+1+\rho/2}(z)}{(2n+1)(2n+5)}.
\end{array}
\end{equation}
Since the series series in (\ref{A24}) are uniformly convergent in $z$ the
limit $z\gg 1$ can be interchanged with the sums. The leading term
is $O(z^{-1})$, independent of $\rho$. Therefore it makes no contribution
to (\ref{A7}). The subleading terms are $O(z^{-2})$ and $\rho$-dependent, but the remainder, formally $O(z^{-3})$, is logarithmically divergent. Hence,
the series in (\ref{A24}) has to be summed before taking the $z\gg 1$ limit 
in order to get the required subleading terms that give $C$ in Eq.(\ref{13}).
The constant $C$ can only originate from (\ref{A24}) through
$\partial f(z,\rho)/\partial\rho\biggr|_{\rho=0}$.
We now give a demonstration of how $C$ arises.

Replace the Bessel functions in (\ref{A24}) with integrals using

\begin{equation}\label{A25}
\int\limits_0^zdx\,xJ_\mu^2(x)=\frac12z^2\left[J_\mu^2(z)-J_{\mu+1}(z)J_{\mu-1}(z)\right].
\end{equation}
The integrals and sums can again be interchanged for all finite $z$.
Then (\ref{A24}) becomes

\begin{equation}\label{A26}
\partial\left[z^2f(z,\rho)\right]/\partial z=\frac{\pi}{2}z\sum\limits_{n=1}^{\infty}\frac{J_{n+2+\rho/2}^2(z)+J_{n+\rho/2}^2(z)}{(2n+1)(n+1)}+4\pi\sum\limits_{n=0}^{\infty}\frac{J_{n+2+\rho/2}^2(z)}{(2n+1)(2n+5)}.
\end{equation}
As required by (\ref{A7}) the series can be differentiated term-by-term
with respect to $\rho$ as $J_\mu(z)$ is an entire function of $\mu$. The limit
$z\gg 1$ can be interchanged with the sums, giving

\begin{equation}\label{A27}
\frac{\partial}{\partial z}\left[z^2\partial f(z,\rho)/\partial\rho\biggl|_{\rho=0}\right]=\left(\frac{\pi}{2}+\frac{\pi}{2}\ln 2-\frac{\pi^2}{4}\right)\cos(2z)-\frac{\pi}{30}\cos(2z)+O\left(\frac{\cos(2z)}{z},\,\frac{\sin(2z)}{z}\right),
\end{equation}
where first (second) term in (\ref{A27}) comes from the first two (last)
series in (\ref{A26}); the indicated remainder has been verified to be
finite. Hence, for $z\gg 1$

\begin{equation}\label{A28}
\partial f(z,\rho)/\partial\rho\biggl|_{\rho=0}=\left(\frac{\pi}{4}\ln 2+\frac{7\pi}{30}-\frac{\pi^2}{8}\right)\frac{\sin(2z)}{z^2}+\frac{D}{z^2}+O\left(\frac{\cos(2z)}{z^3},\,\frac{\sin(2z)}{z^3}\right),
\end{equation}
where $D$ is the constant of integration of (\ref{A27}). In Eq.(\ref{13}),
$C=8D/3$. The reason $D$ is hard to calculate from (\ref{A24}) is
because it arises in the transitional region where $J_\mu(z)$ has
$\mu\sim z\gg 1$. Of course we cannot {\it a priori} rule out that $C$ is zero
in our way of calculating the infrared divergent part of $\partial\Pi_2/\partial m^2$.

The remaining terms in (\ref{A7}) are tedious but straightforward
to calculate. The result for $z=kL\gg 1$ is given by Eq.(\ref{13}).

Finally, the terms $O(\nu^4)$ and higher in $S_2$ can be calculated
by continuing the expansion of $\sigma$ in powers of $\nu$, which we noted
converges when $\nu\le 2$ and $j\ge 1/2$. This will involve higher derivatives
in $\rho$ of the second-order result together with new terms that are
easily estimated. Each higher order only contains terms
$O\left(\displaystyle\frac{\cos(2kL)}{(kL)^2},\,\frac{\sin(2kL)}{(kL)^2}\right)$; there are no terms of the form $C/(kL)^2$.
The case $\nu>2$ is dealt with in Appendix B.

\vskip1cm

\begin{center}
APPENDIX B
\end{center}

\setcounter{equation}{0}

\renewcommand{\theequation}{B\arabic{equation}}

The purpose of this appendix is to continue the results of
Appendix A to values of $\nu>2$. Recall that the first zero mode
in the positive chirality sector occurs when $\nu>2$.

The estimates below require the following results: From

\begin{equation}\label{B1}
\begin{array}{cc}
\sigma=\left[(2j+1)^2+4M\nu+\nu^2\right]^{1/2},\quad -j+1/2\le M\le j-1/2, \\
\sigma\ge\sqrt{8j},\quad j=\frac12,1,\ldots,
\end{array}
\end{equation}
for all real values of $\nu$. Also

\begin{equation}\label{B2}
\frac{4M\nu+\nu^2}{(2j+1)^2}\le 1,\ j\ge 1/2,\ \nu\le 2;\ j\ge j_*,\ \nu>2,
\end{equation}
where

\begin{equation}\label{B3}
j_*=\left\{\frac{\nu-1}{2}+\sqrt{\frac12\nu^2-\nu}\right\},
\end{equation}
and $\{x\}$ is the greatest integer/half-integer value of $x$ exceeding
or equal to $x$. Therefore the expansion of $\sigma$ in powers of $\nu$ converges for $j\ge j_*$ for $\nu>2$ and for $j\ge 1/2$ for $\nu\le 2$.

Consider the first term in $S_2$  given by Eq.(\ref{11}). For all $\nu>0$
it can be rewritten as

\begin{equation}\label{B4}
S_{21}(\nu, z)=\sum\limits_{j=\frac12}^{j_*-\frac12}(2j+1)\sum\limits_{M=-j+\frac12}^{j-\frac12}J_{\sigma+1}^{2}(z)+\sum\limits_{j=j_*}^{\infty}(2j+1)\sum\limits_{M=-j+\frac12}^{j-\frac12}J_{\sigma+1}^2(z),
\end{equation}
with $z=kL$. Note that for real $z$

\begin{equation}\label{B5}
|J_\sigma(z)|\le\frac{|\frac12z|^\sigma}{\Gamma(\sigma+1)},\quad\sigma\ge-\frac12.
\end{equation}
From (\ref{B1}) and $\sigma\le\sqrt2(2j+1)$ from (\ref{B2}), it follows from (\ref{B5}) with $z=kL>2$ that there is a $\nu$ - independent upper bound

\begin{equation}\label{B6}
|J_\sigma(z)|\le\frac{(\frac12z)^{\sqrt2(2j+1)}}{\Gamma(\sqrt{8j}+1)},\quad j\ge\frac12,\ \nu\le 2;\ j\ge j_*,\ \nu>2,
\end{equation}
on the terms in the second series in (\ref{B4}). The sum of the bounded
terms is absolutely convergent, and so the second series in (\ref{B4})
is uniformly convergent for all $\nu>0$.

Now $J_{\sigma+1}$ is an entire function of $\sigma$, and hence so is $J_{\sigma+1}^2$. Therefore the expansion of $J_{\sigma+1}^2$ about the point $\sigma=2j+1$ has an infinite radius of convergence for any $\nu$:

\begin{equation}\label{B7}
J_{\sigma+1}^2(z)=\sum\limits_{n=0}^{\infty}J_{2j+2}^{2(n)}(z)\frac{(\sigma-2j-1)^n}{n!},
\end{equation}
where $J_{2j+2}^{2(n)}=d^nJ_\mu^2/d\mu^n\biggl|_{\mu=2j+2}$. Because of the
uniform convergence of the second series in (\ref{B4}) the sums over
$n$ and $j$ can be interchanged when (\ref{B7}) is inserted in (\ref{B4}):

\begin{equation}\label{B8}
S_{21}(\nu,z)=\sum\limits_{j=\frac12}^{j_*-\frac12}(2j+1)\sum\limits_MJ_{\sigma+1}^2(z)+\sum\limits_{n=1}^{\infty}S_n(\nu,z)/n!,
\end{equation}
where

\begin{equation}\label{B9}
S_n(\nu,z)=\sum\limits_{j=j_*}^{\infty}(2j+1)J_{2j+2}^{2(n)}(z)\sum\limits_M(\sigma-2j-1)^n.
\end{equation}
Note that

\begin{equation}\label{B10}
(\sigma-2j-1)^n=\left[\frac12\,\frac{4M\nu+\nu^2}{2j+1}-\frac18\,\frac{(4M\nu+\nu^2)^2}{(2j+1)^3}+\ldots\right]^n=2^{-n}\left(\frac{4M\nu+\nu^2}{2j+1}\right)^n(1+R)^n,
\end{equation}
where, because the series is convergent and alternating,

\begin{equation}\label{B11}
O<|R|<\frac14\,\frac{4M\nu+\nu^2}{(2j+1)^2}\le\frac14,\quad j\ge j_*.
\end{equation}

Hence,

\begin{equation}\label{B12}
\sum\limits_{n=1}^{\infty}S_n(\nu,z)/n!=\sum\limits_{n=1}^{\infty}\frac{1}{2^nn!}\sum\limits_{j=j_*}^{\infty}(2j+1)J_{2j+2}^{2(n)}(z)\sum\limits_M\left(\frac{4M\nu+\nu^2}{2j+1}\right)^n(1+R)^n.
\end{equation}
For $j\gg j_*$ in (\ref{B12}) the leading term is

\begin{equation}\label{B13}
\sum=2\sum\limits_{n=1}^{\infty}\frac{\nu^n}{n!}\sum\limits_{j\gg j_*}^{\infty}j^{1-n}J_{2j+2}^{2(n)}(z)\sum\limits_MM^n,
\end{equation}
where $R=O(\nu/j)$ has been neglected. From the estimate

\begin{equation}\label{B14}
\sum\limits_{M=-j+\frac12}^{j-\frac12}M^n=\frac{2}{n+1}\left(j-\frac12\right)^{n+1},\quad n=2,4,\ldots,
\end{equation}

and letting $n\to 2n$ one gets

\begin{equation}\label{B15}
\sum=4\sum\limits_{n=1}^{\infty}\frac{\nu^{2n}}{(2n+1)(2n)!}\sum\limits_{j\gg j_*}^{\infty}j^2J_{2j+2}^{2(2n)}(z).
\end{equation}
The convergence of the series in (\ref{B15}) is investigated by estimating
the sum over $j$ by the integral

$$
\int\limits_{Nj_*}^{\infty}dj\,j^2J_{2j+2}^{2(2n)}(z),
$$
for $N,n\gg 1$ and employing the method of steepest descents for fixed
$z$ using

\begin{equation}\label{B16}
J_\mu(z){\phantom{abcd}\atop\widetilde{\stackrel{\quad}{\scriptstyle \mu\gg 1}}}\frac{1}{\sqrt{2\pi\mu}}\left(\frac{ez}{2\mu}\right)^\mu. 
\end{equation}
This gives

\begin{equation}\label{B17}
\sum\limits_{j\gg j_*}j^2J_{2j+2}^{2(2n)}(z){\phantom{abcd}\atop\widetilde{\stackrel{\quad}{\scriptstyle n\gg 1}}}\frac18\left(\frac{n}{\pi(\ln n)^3}\right)^{1/2}e^{2n\ln\,\ln n+O\left(n\ln\,\ln n/\ln n\right)}. 
\end{equation}
In view of (\ref{B17}) it is evident that the power series in $\nu$ in
(\ref{B15}) converges for all $\nu>0$ satisfying (\ref{B2}), (\ref{B3}) and for all $z>0$, and hence so does the series $S_{21}$ in (\ref{B8}).

Repeat the above analysis for the remaining terms in $S_2$ in
Eq.(\ref{11}). With obvious notation,

\begin{equation}\label{B18}
\begin{array}{cc}
\displaystyle S_2(\nu,z)=\sum\limits_{j=\frac12}^{j_*-\frac12}(2j+1)\sum\limits_M\left[J_{\sigma+1}^2-J_{\sigma+2}J_{\sigma}+J_{\sigma-1}^2-J_{\sigma-2}J_\sigma-(\nu=0)\right] \\
\hskip-1cm\displaystyle+\sum\limits_{n=1}^{\infty}\sum\limits_{j=j_*}^{\infty}(2j+1)\sum\limits_M[\,\bullet\,]_{\sigma=2j+1}^{(n)}(\sigma-2j-1)^n/n!.
\end{array}
\end{equation}
Denote the first sum in (\ref{B18}) by $\sum_1$. Setting $z=kL$ then
\begin{equation}\label{B19}
{\sum}_1\ {\phantom{abcd}\atop{{=\hskip-0,36cm=}\atop{\scriptstyle{kL\gg 1}}}}\ \sum\limits_{j=\frac12}^{j_*-\frac12}(2j+1)\sum\limits_{M=-j+\frac12}^{j-\frac12}\left(\frac{2\cos(2kL-\pi\sigma)}{\pi(kL)^2}+O(kL)^{-3}\right).
\end{equation}

We know from our analysis of $S_{21}$ that the second series in (\ref{B18}) is
absolutely convergent, and so it can be rearranged in powers of $\nu$.
Due to the symmetric sum over $M$ only even powers of $\nu$ survive.
Denoting the second series by $\sum_2$, it takes the form
\begin{equation}\label{B20}
{\sum}_2=\sum\limits_{n=1}^{\infty}\frac{\nu^{2n}}{(2n)!}\sum\limits_{j=j_*}^{\infty}(2j+1)\sum\limits_{M=-j+\frac12}^{j-\frac12}a_{2n}(j,kL).
\end{equation}
In second order we found for $kL\gg 1,\ \nu\le 2$,
\begin{equation}\label{B21}
\frac{\nu^2}{2!}\sum\limits_{j=\frac12}^{\infty}(2j+1)\sum\limits_Ma_2(j,kL)=-\frac{2\nu^2}{\pi kL}+\frac{C\nu^2}{(kL)^2}+O\left(\frac{\nu^2\cos(2kL)}{(kL)^2},\,\frac{\nu^2\sin(2kL)}{(kL)^2}\right).
\end{equation}
In fourth and higher orders following Appendix A only terms of
$O\left(\displaystyle\frac{\cos(2kL)}{(kL)^2},\frac{\sin(2kL)}{(kL)^2}\right)$ are present. Therefore set for $n\ge 2,\, kL\gg 1$
\begin{equation}\label{B22}
\sum\limits_{j=\frac12}^{\infty}(2j+1)\sum\limits_{M=-j+\frac12}^{j-\frac12}a_{2n}(j,kL)=b_{2n}\frac{\sin(2kL)}{(kL)^2}+c_{2n}\frac{\cos(2kL)}{(kL)^2}+O\left((kL)^{-3}\right).
\end{equation}
Then for $kL\gg 1$
\begin{equation}\label{B23}
{\sum}_2=-\frac{2\nu^2}{\pi kL}+\frac{C\nu^2}{(kL)^2}+\sum\limits_{n=1}^{\infty}\frac{\nu^{2n}}{(2n)!}b_{2n}\frac{\sin(2kL)}{(kL)^2}+\sum\limits_{n=1}^{\infty}\frac{\nu^{2n}}{(2n)!}c_{2n}\frac{\cos(2kL)}{(kL)^2}+O\left((kL)^{-3}\right).
\end{equation}
Since $\sum_2$ is convergent for $\nu\le 2$ with $j_*=1/2$ and sine and cosine are linearly independent, both series in (\ref{B23}) are convergent as well.
For $\nu>2$ the sums in (\ref{B20})-(\ref{B22}) begin at $j=j_*\ge1$ given by (\ref{B3}), 
and 
following the 
above analysis they are convergent. Terms with $j=1/2,..,j_*-1/2$
result in the oscillating terms in (\ref{B19}) whose power series
expansion in $\nu$ is not guaranteed to converge for $\nu>2$. Therefore,
the results (\ref{B21}), (\ref{B22}), as well as (\ref{B23}) and Eq.(\ref{13}) continue to hold for $\nu>2$, except that the coefficients of $\cos(2kL)/(kL)^2$
and $\sin(kl)/(kL)^2$ will change. This result is consistent with
what we found for the fully summed series for $S_1$  in Eq.(\ref{12}).

\end{document}